# Redefining Data-Centric Design: A New Approach with a Domain Model and Core Data Ontology for Computational Systems

William Johnson, James Davis, Tara Kelly,

William.johnson.22@gmail.com, James.davis34@gmail.com, Tarakelly963@gmail.com,

**Abstract:** This paper presents an innovative data-centric paradigm for designing computational systems by introducing a new informatics domain model. The proposed model moves away from the conventional node-centric framework and focuses on data-centric categorization, using a multimodal approach that incorporates objects, events, concepts, and actions. By drawing on interdisciplinary research and establishing a foundational ontology based on these core elements, the model promotes semantic consistency and secure data handling across distributed ecosystems. We also explore the implementation of this model as an OWL 2 ontology, discuss its potential applications, and outline its scalability and future directions for research. This work aims to serve as a foundational guide for system designers and data architects in developing more secure, interoperable, and scalable data systems.

## 1. Introduction

The creation of the Internet on January 1, 1983, marked a transformative shift in global connectivity, establishing a new era for communication networks. Before this, fragmented computer networks struggled to communicate seamlessly. The introduction of the Transmission Control Protocol/Internet Protocol (TCP/IP) enabled consistent data transfer and became the standard for digital communication. However, this node-centric approach, which relies heavily on Internet Protocol (IP) addresses, has also created significant security vulnerabilities and privacy concerns due to its focus on network nodes rather than the data itself.

In today's digital landscape, the centralized aggregation and storage of sensitive user data — including IP addresses — by service providers pose substantial security risks. These centralized repositories are prime targets for cyberattacks, potentially compromising user privacy and exposing sensitive information. Additionally, the reliance on IP-based system modeling has amplified these risks, necessitating a shift toward a more secure and resilient design approach.

This paper proposes a novel data-centric design methodology that moves away from traditional node-focused models. By prioritizing data as the central entity and incorporating multimodal frameworks encompassing objects, events, concepts, and actions, this approach enhances data security and flexibility. The new informatics domain model reimagines data's role in system design, emphasizing its importance throughout its entire lifecycle to foster innovation, security, and seamless data interoperability.

## 2. Problem Statement

Current computational systems are challenged by outdated informatics models that inadequately prioritize data semantics. Traditional node-centric models, which focus on securing IP addresses during data transmission, often treat data semantics as an afterthought, addressed only during data curation. This approach frequently leads to inefficient schema design, resulting in poorly structured data that undermines security and interoperability.

One major challenge in existing models relates to the cryptographic principles of data "integrity" and "authenticity." While both are essential for ensuring data reliability during transmission and storage, they are often not maintained simultaneously, leading to security weaknesses. For example, ensuring the authenticity of a data source does not always guarantee the integrity of the associated metadata or content, complicating the preservation of the original context throughout the data lifecycle.

The current data landscape also faces issues related to data quality, particularly with unstructured data, which constitutes around 80% of recorded data. Unlike structured data tied to metadata, unstructured data is less accessible for analysis by artificial intelligence, machine learning, and statistical tools. As a result, there is a pressing need for a new model that prioritizes data semantics from the outset to ensure data is consistently interpretable, secure, and meaningful across various digital ecosystems.

## 3. Objectives

This paper presents a comprehensive solution based on a data-centric design framework to address the limitations of current informatics models. The proposed model aims to achieve three primary objectives:

Enhancing Security through a Data-Centric Model: By categorizing data into four distinct modalities — objects, events, concepts, and actions — the model facilitates secure, role-based access tailored to specific user roles and responsibilities. This categorization helps mitigate unauthorized access risks and data breaches, ensuring data is accessible only to those with appropriate privileges.

Developing a Core Data Ontology: A key goal is to build a granular, flexible core data ontology that systematically classifies data components within the model's framework. The ontology provides a consistent reference for data architects and system designers, promoting coherent knowledge representation across distributed ecosystems.

Establishing a Standard Framework for Informatics Design: The model aims to become a foundational reference for designing secure and interoperable data systems. It offers a unifying framework that transcends traditional boundaries, supporting effective data governance and facilitating collaboration among stakeholders in diverse domains.

## 4. Methodology

The development of the informatics domain model and its core data ontology was grounded in a multidisciplinary approach, incorporating insights from data science, linguistics, mathematics, philosophy, and other fields. The initial research, which began in mid-2020, focused on understanding the dual aspects of data semantics and machine assurance, using precise terminology from authoritative sources such as the Oxford English Dictionary.

The methodology involved a bottom-up design process, free from legacy model constraints, allowing the model to evolve organically based on robust, well-defined principles. Subsequent phases of the research included developing the core ontology in 2023, which provides a flexible framework for dynamic data categorization and search within distributed systems.

By adopting this rigorous, interdisciplinary approach, the model provides a versatile foundation for understanding and managing the complexities of data-centric design in computational systems.

## 5. Applications and Benefits

The proposed data-centric design model and its accompanying core data ontology offer numerous advantages across a wide range of domains and industries. By shifting the focus from node-centric approaches to a comprehensive data-centric paradigm, organizations can achieve more secure, efficient, and interoperable data management. The following outlines the key applications and benefits:

## 5.1 Enhanced Data Management and Governance

The data-centric model provides a robust foundation for data management and governance by structuring data around four core modalities: objects, events, concepts, and actions. This approach enables organizations to gain a holistic view of their data assets and develop strong governance frameworks that enhance data quality, integrity, and consistency across the entire data lifecycle.

Application Example: In the financial services industry, a bank can utilize the model to better manage customer information. By categorizing data into objects (such as customer details), events (like transactions or interactions), concepts (risk categories, customer segmentation), and actions (fraud detection algorithms), the bank can maintain accurate and secure data management. This approach also supports compliance with stringent regulatory requirements, improves customer experience, and enhances decision-making through more reliable data analytics.

## 5.2 Facilitating Semantic Interoperability

Semantic interoperability is the ability of different systems, applications, or organizations to exchange and use data seamlessly, without loss of meaning. The data-centric model promotes semantic interoperability by providing a unified and standardized representation of data. The core data ontology, with its well-defined relationships and classifications, enables different systems to interpret and integrate data consistently.

Application Example: In healthcare, achieving semantic interoperability is critical for efficient data exchange among hospitals, clinics, and research institutions. By adopting a common ontology and understanding of data semantics, patient records, diagnostic data, treatment histories, and research findings can be seamlessly shared and understood across different platforms. This reduces medical errors, enhances patient care, enables coordinated treatment plans, and accelerates research and innovation in healthcare.

### 5.3 Intelligent Decision-Making and Automation

The model enhances intelligent decision-making and automation by incorporating computational intelligence and action-based functionalities. By leveraging a structured representation of data, organizations can build systems that analyze and interpret data in real-time, enabling automated decision-making, predictive analytics, and optimization.

Application Example: In manufacturing, the data-centric model can optimize production processes. By capturing events related to equipment performance (such as maintenance activities or failures), monitoring objects like machine components, defining concepts for quality control, and enabling automated actions (like real-time adjustments to machinery), manufacturers can reduce downtime, improve product quality, and enhance overall efficiency. Predictive maintenance algorithms, powered by this structured data, can foresee potential equipment failures, reducing costs and improving productivity.

### 5.4 Improved Knowledge Management and Discovery

Knowledge management is the process of capturing, distributing, and effectively using organizational knowledge. The data-centric model facilitates this by providing a structured framework for organizing and categorizing knowledge assets. The ontology's concepts and relationships enable the creation of knowledge graphs, semantic networks, and intelligent search systems, thereby enhancing knowledge discovery and sharing.

Application Example: A research institution can use the model to manage its vast collection of publications, datasets, and other knowledge assets. By organizing these assets into structured domains, such as concepts (thematic areas of research), events (research activities or milestones), objects (datasets or experimental results), and actions (analysis methods), the institution can create a powerful knowledge discovery platform. Researchers can navigate this platform to find related work, discover new insights, and collaborate more effectively, leading to accelerated innovation and knowledge creation.

### 5.5 Strengthening Data Security and Privacy

A core advantage of the data-centric model is its ability to enhance data security and privacy through role-based access control. By categorizing data into discrete modalities, the model allows for granular control over who has access to specific types of data, based on their roles and responsibilities.

Application Example: In the context of government and public administration, data security and privacy are paramount. Using the data-centric model, government agencies can implement secure, role-based access to sensitive information. For instance, access to citizens' personal data (objects) might be restricted to authorized personnel, while aggregate statistical data (events) could be made publicly available. This approach minimizes the risk of unauthorized access, data breaches, and misuse of sensitive information, while still enabling transparency and data sharing where appropriate.

### 5.6 Supporting Data-Driven AI and Machine Learning

The model provides a strong foundation for developing data-driven artificial intelligence (AI) and machine learning (ML) applications. By structuring data into four distinct domains, the model facilitates data preprocessing, feature engineering, and normalization processes critical for training accurate and robust AI/ML models. The granular representation of objects, events, concepts, and actions allows for better feature extraction, contextual understanding, and predictive modeling.

Application Example: A retail company can use the data-centric model to build personalized recommendation engines. By analyzing customer purchase history (events), understanding product attributes and categories (objects), inferring customer preferences and buying patterns (concepts), and automating recommendation algorithms (actions), the company can deliver tailored product suggestions to individual customers. This enhances the customer experience, increases sales, and builds brand loyalty by providing value through personalized engagement.

### 5.7 Enhancing Collaboration Across Diverse Systems and Domains

The data-centric model's ontology fosters communication, collaboration, and standardization by providing a common framework for understanding and categorizing data. This unifying framework is particularly valuable in environments with multiple stakeholders and disparate data systems, such as cross-industry partnerships, supply chains, and multi-organizational projects.

Application Example: In the field of smart cities, data from various sources — including traffic sensors, public transportation systems, weather stations, and emergency services — must be integrated and analyzed to make urban living more efficient and sustainable. The data-centric model enables diverse systems to communicate effectively by adopting a shared ontology, allowing city planners, engineers, and public officials to collaborate using a unified data framework. This leads to better decision-making, more responsive public services, and an improved quality of life for city residents.

**5.8 Promoting Scalability and Adaptability in System Design**

The data-centric model is designed to be scalable and adaptable, capable of integrating with existing and future ontologies, datasets, and vocabularies. Its flexible structure allows it to be applied across various domains and industries, making it suitable for a wide range of applications from small-scale systems to large, complex infrastructures.

Application Example: In the energy sector, where the need for scalable data management is critical due to the vast amounts of data generated from various sources like smart meters, grid sensors, and power plants, the data-centric model offers a scalable solution. It allows energy companies to manage, analyze, and optimize data flow efficiently, supporting grid reliability, demand forecasting, and sustainable energy practices. As the energy landscape evolves with new technologies and regulatory requirements, the model's adaptability ensures it remains relevant and effective.

**Conclusion of Benefits**

Overall, the data-centric design model and its core data ontology provide a versatile, secure, and efficient framework that can be leveraged across diverse fields. From data management and governance to enhancing collaboration and supporting AI and machine learning applications, this model lays the groundwork for organizations to unlock the full potential of their data assets. It not only enhances operational efficiency and security but also paves the way for innovation and strategic growth in an increasingly data-driven world.

**6. Conclusion**

This paper introduces a transformative data-centric model for computational systems, emphasizing a multimodal approach that prioritizes data over nodes. By shifting the focus from IP-based security to data-centric design, the model offers a more secure, flexible, and scalable framework for the digital age. Future research will explore its applications in diverse fields, such as healthcare, finance, and education, and its integration with emerging technologies like AI and IoT.